\documentclass[epjCONF, onecolumn]{svjour}
\usepackage{graphics}
\usepackage{epsfig}
\usepackage[varg]{txfonts} 
\usepackage[latin1]{inputenc}
\session-title{Assembling the Puzzle of the Milky Way}
\begin{document}
\title{Modeling disc non-axisymmetries: multiple patterns, radial migration, and thick discs}

\author{Ivan Minchev \inst{1}\fnmsep\thanks{\email{ iminchev1@gmail.com}} \and Benoit Famaey\inst{2,3} \and Alice C. Quillen \inst{4} \and Walter Dehnen \inst{5}
}
\institute{Leibniz-Institut f\"{ur} Astrophysik Potsdam (AIP), An der Sternwarte 16, D-14482, Potsdam, Germany 
\and Observatoire Astronomique de Strasbourg, CNRS UMR 7550, France 
\and Argelander Institute for Astronomy, Bonn, Germany 
\and Department of Physics and Astronomy, University of Rochester, Rochester, NY 14627
\and Theoretical Astrophysics Group, Department of Physics \& Astronomy, University of Leicester, Leicester LE1 7RH}
\abstract{
Disc non-axisymmetrc components, such as spirals and central bars, are nowadays known to play an important role in shaping galactic discs. Here we use Tree-SPH N-body simulations to examine the effect of these perturbers on two aspects: the occurrence of multiple patterns in discs and the effects of radial migration on disc thickening. We find that, in addition to a central bar, multiple spiral patterns and lopsided modes develop in all models. Interaction among these asymmetric features results in a large scale stellar migration. However, we show that, despite the strong radial mixing, discs cannot be thickened sufficiently to match observed thick discs. We relate this to the adiabatic cooling as stars migrate radially outwards. We also find that the bulge contribution to a thick-disc component for an Sa-type galaxy at $\sim$~2.5 disc scale-lengths is less than 1\% and zero in the case of a Milky Way-like, Sb-type. 
Our findings cast doubt on the plausibility of thick disc formation via stellar radial migration.
} 
\maketitle
\section{Introduction}
\label{sec:intro}

Major progress has been made over the last few years in understanding disc formation in cosmological simulations, although many discrepancies remain between the predictions of such simulations and the properties of the disc galaxies we observe today (e.g., \cite{scannapieco09}). Furthermore, there is now increasing evidence that mergers may not necessarily be the dominant process in the formation of discs, but internal evolution processes also play a major role in shaping galactic discs (e.g., \cite{bournaud09,minchev11}. Recent theoretical studies have found that: (i) extended discs can be explained by satellite- (\cite{quillen09}) or spiral-driven migration (\cite{roskar08}); (ii) angular momentum bar-halo coupling (\cite{athanassoula07}) and a powerful radial migration mechanism involving bar-spiral arms interaction (\cite{mf10,minchev11}) are at play in barred spirals, such as the Milky Way; and (iii) thick discs, traditionally explained by heating due to mergers and/or accretion of satellites (\cite{meza05}), could also be caused by secular evolution linked to gas-rich, turbulent, clumpy discs (\cite{bournaud09}) or radial migration (\cite{schonrich09,loebman11}). 

It is, however, crucial to have a clear understanding of how much can be expected from secular disc evolution. In particular, here we would like to examine in detail the effect of radial migration on the formation of thick discs. To this end, we study three isolated giant galaxies from the GalMer database\footnote{The GalMer database is accessible online at http://galmer.obspm.fr/} (\cite{dimatteo07,chilingarian10}): the gSb (20\% gas, $v_c=220$~km/s), gSa (10\% gas, $v_c=280$~km/s), and gS0 (Identical to gSa but no gas) models. In these Tree-SPH N-body simulations the gSb model develops a bar similar to the one seen in the Milky Way (MW) (\cite{minchev07,minchev10}), while both the gS0 and gSa bars are substantially larger than that (see Fig.~\ref{fig:2}). We have shown (\cite{minchev11}) that this has strong effects on the migration efficiency found in these discs, with larger bars giving rise to stronger mixing (also seen in studies of diffusion coefficients, e.g., \cite{brunetti11,shevchenko11}). 

\section{Multiple patterns in galactic discs}
\label{sec:multiple}

A number of previous works have seen spiral structure in N-body or particle-mesh simulations at pattern speeds different from the bar's (e.g., \cite{sellwood88,patsis99}, see also \cite{famaey11} for observational evidence in the MW). It has been suggested that these multiple waves are coupled (\cite{rautiainen99,masset97}) and thus, longer-lived ($>0.5$~Gyr,\cite{quillen11}), rather than transient (on the order of a dynamical time).  

\begin{figure}
\resizebox{1.0\columnwidth}{!}{%
\includegraphics{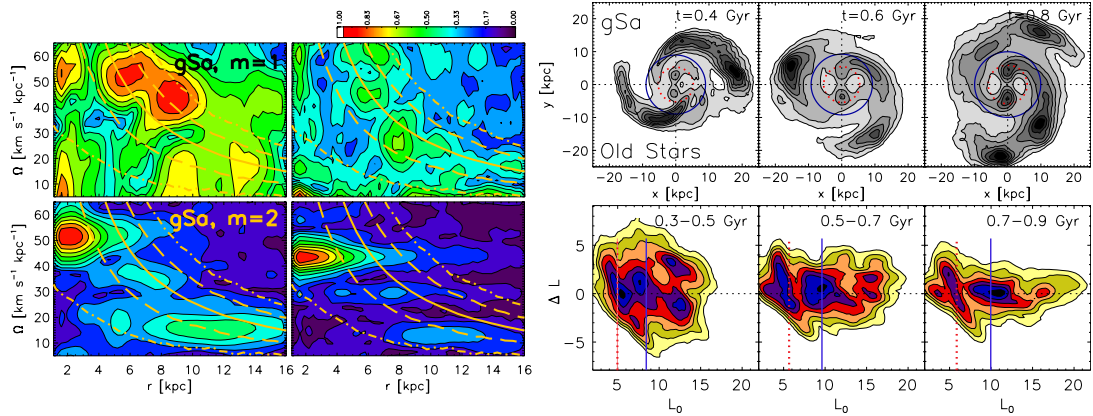} }
\caption{
{\bf First and second columns:} power spectrograms of the $m=1$ (top) and m=2 (bottom) Fourier components for the gSa model. The first and second columns show time windows in the range $t=[0.3-0.9]$~Gyr and $t=[0.8-1.3]$~Gyr, respectively. {\bf Columns 3-5, top row:} the differential density of the gSa disc for three different times, as indicated in each panel. The bars are aligned with the vertical axis and their CR and OLR radii are shown by the dotted red and solid blue circles, respectively. {\bf Columns 3-5, bottom row:} present the incremental changes in angular momentum, $\Delta L$, in time intervals of 200~Myr, centered on the time of the snapshots shown right above. Both axes are divided by the estimated rotation curve for the galaxy, thus the values displayed are approximately equal to galactic radius in kpc. Note that discontinuities in the spiral arms in the outer disc correspond to strong increases in $\Delta L$.
}
\label{fig:1}       
\end{figure}

The simulations used in this work present yet another example of the propagation of multiple waves in galactic discs. In the first two columns of Fig.~\ref{fig:1} we plot Fourier power spectrograms for the $m=1$ (top, lopsided structure) and $m=2$ (bottom, bar or 2-armed spiral structure) components for the gSa model. The time windows used for earlier (first column) and later (second column) times are $t=[0.3-0.9]$~Gyr and $t=[0.8-1.3]$~Gyr, respectively. The bar is identifiable by the strongest feature in the inner disc (0-6 kpc, bottom panels); two additional features (spirals) are present in the outer disc at lower pattern speeds. The similarity between the Fourier spectrograms at different stages in the simulations (as the bar slows down) suggest that our discs also exhibits couplings between waves, with the bar possibly driving the spiral waves.

To see how these multiple patterns appear in the disc morphology, in columns 3-5, top row of Fig.~\ref{fig:1} we plot the face-on stellar density of the gSa model, with the azimuthally averaged component subtracted. As can be seen here (but also in all other models), all time snapshots show discontinuities/breaks (strong decrease in amplitude) in the spirals at the bar's outer Lindblad resonance OLR, as well as outside the OLR. Discontinuities in spiral arms can indicate changes in the dominant pattern or constructive/destructive interference between different spiral modes. 
For example, the lack of rotational symmetry in the snapshots in Fig.~\ref{fig:1} is indicative of an $m=1$ mode, possibly resulting from interference between two patterns of different multiplicity (e.g., a bar and a three-armed spiral wave). Strong $m=1$ modes are found in the power spectrograms as well. Observationally, lopsidedness has been shown to be a common property of spiral galaxies. For example, \cite{zaritsky97} have shown that $\sim$~30\% of field spiral galaxies in a magnitude-limited sample exhibit significant lopsidedness in the outer discs.
 
 \subsection{Angular momentum exchange at spiral discontinuities}
\label{sec:am}

We now study the angular momentum exchange at different radii in the gSa disc. In Fig.~2 by \cite{minchev11} we presented the cumulative changes in angular momentum for this same model, as a function of radius for the duration of the simulation (3~Gyr). In contrast, the bottom panels of rows 3-5 of Fig.~\ref{fig:1} show the {\it incremental} changes in the time intervals 0.3-0.5, 0.5-0.7, and 0.7-0.9~Gyr. We estimate these as $\Delta L(r)=L_1(r)-L_0(r)$, where $L_1(r)$ and $L_0(r)$ are the final and "initial" (i.e., t=0.3, 0.5, and 0.7) angular momenta at each time step. In other words, each plot shows how much migration has occurred as a function of radius during the time step of 200~Myr, centered on the time of the snapshots shown in the first three columns of the same figure. Both axes are divided by the estimated rotation curve, thus the values displayed are approximately equal to galactic radius in kpc. The bar's corotation (CR) and OLR are shown as the dotted and solid vertical lines in each plot. 
Examining the changes of angular momentum, we note that the effect near the bar's OLR is similar to that at the CR and even stronger in the first time period shown. Very well defined multiple peaks in $\Delta L$ are present at all time steps, with strongest features near the spiral discontinuities. 

The relationship we find here between the changes in angular momentum and the gaps and breaks in the spiral structure occurring at the same radii, makes it compelling to interpret these with resonance overlap regions associated with the multiple patterns. Beyond the outermost break however, spiral features are most likely not self-gravitating because of the low densities in those regions. This is also suggested by the abrupt decrease in pitch angle. We nevertheless see changes at those discontinuities. These must be related to gravity torques \cite{foyle10}, rather than resonance overlap, given the long dynamical times at those radii.

\section{Radial migration and thick discs}

\begin{figure}
\resizebox{0.95\columnwidth}{!}{%
\includegraphics{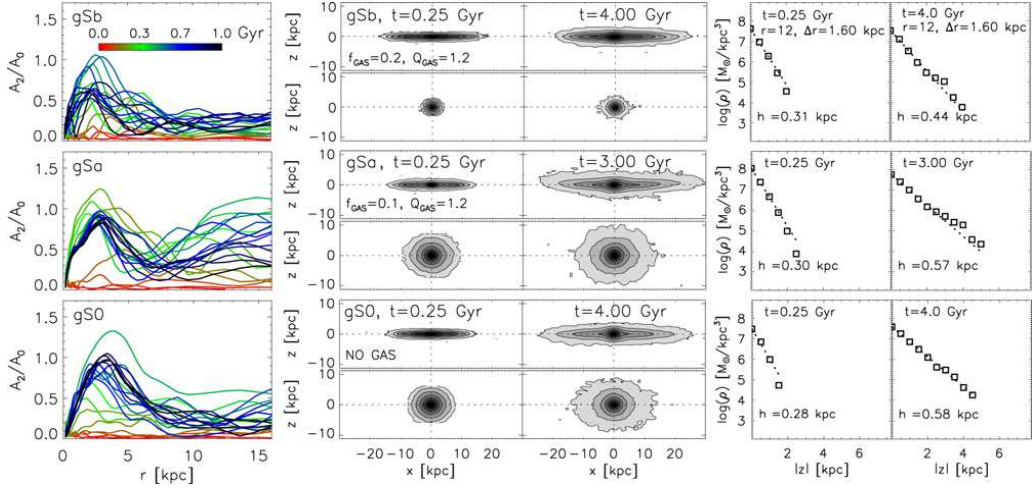} }
\caption{
{\bf First column:} m=2 Fourier amplitudes, $A_2$, as a function of time (color bar) for our model galaxies. {\bf Columns 2-3:} edge-on view for each disc and bulge (separately) for earlier and later times of their evolution, as indicated in each panel. {\bf Columns 4-5:} scale-heights, $h$, at $\sim2.5$ disc scale-lengths. Note that in 3-4~Gyr of evolution with strong perturbation from bars and spirals, $h$ only doubles despite the efficient stellar radial migration (see Fig.~\ref{fig:1}).
}
\label{fig:2}
\end{figure}

An interesting question that could help us understand the formation and evolution of disc galaxies is how their thick discs came into existence. Recently there has been a growing conviction that radial migration can result in a thick disc formation by bringing out stars with high velocity dispersion from the inner disc and the bulge. To test this hypothesis, here we analyze the three models above. The gSb model is comparable to the MW in mass, rotation curve, spiral structure strengths, and bar and bulge sizes. The gSa (dissipational) and gS0 (dissipationless) models both develop bars and spirals stronger than those of the MW. Both gSb and gS0 exhibit multiple patterns in their discs, similarly to gSa (see Sec.~\ref{sec:multiple}). As shown in Fig.~\ref{fig:1}, this gives rise to strong radial migration. To compare the three models, in the first column of Fig.~\ref{fig:2} we plot the $m=2$ Fourier amplitudes, $A_2/A_0$, as a function of radius, where $A_0$ is the axisymmetric component and $m$ is the multiplicity of the pattern. These indicate the bi-symmetric structure in the disc. For example, for the gS0 model the bar is identifiable by the smooth curve in the inner disc, which at later times has a maximum at $\sim3$~kpc and drops almost to zero at $\sim8$~kpc. Deviations from zero seen beyond that radius are due to the spiral structure. Note that the spirals are strongest for the gSa model, related to the initial 10\% gas fraction in its disc. The gSb model is the one with parameters closest to the MW: bar size $\sim3-4$~kpc, $v_c\sim220$~km/s, and a small bulge.

To see how much discs thicken, in columns 2-3 of Fig.~\ref{fig:2} we plot the edge-on view for each disc and bulge (separately) for earlier (0.25~Gyr) and later (4~Gyr for gSb and gS0, and 3~Gyr for gSa) times of their evolution, as indicated in each panel. Significant thickening is seen for all models, especially for gSa and gS0 due to their larger bars. The bulges are also found to expand, not significantly so for the gSb. Although from these plots it is evident that the discs thicken, it is important to see how this is reflected in their scale-heights. These are shown in columns 4-5 at a radius $\sim2.5$ disc scale-lengths, which is comparable to the solar distance from the Galactic center. 

Surprisingly, we find that, despite the strong radial migration, the initial scale-heights, $h$, hardly double in the 3-4 Gyr of evolution for gSa and gS0, respectively, and increase by only $\sim$~50\% in 4~Gyr for gSb. We look for the reason for this behavior in the next Section.

\subsection{Cooling of stellar samples migrating outwards}

\begin{figure}
\resizebox{0.95\columnwidth}{!}{%
\includegraphics{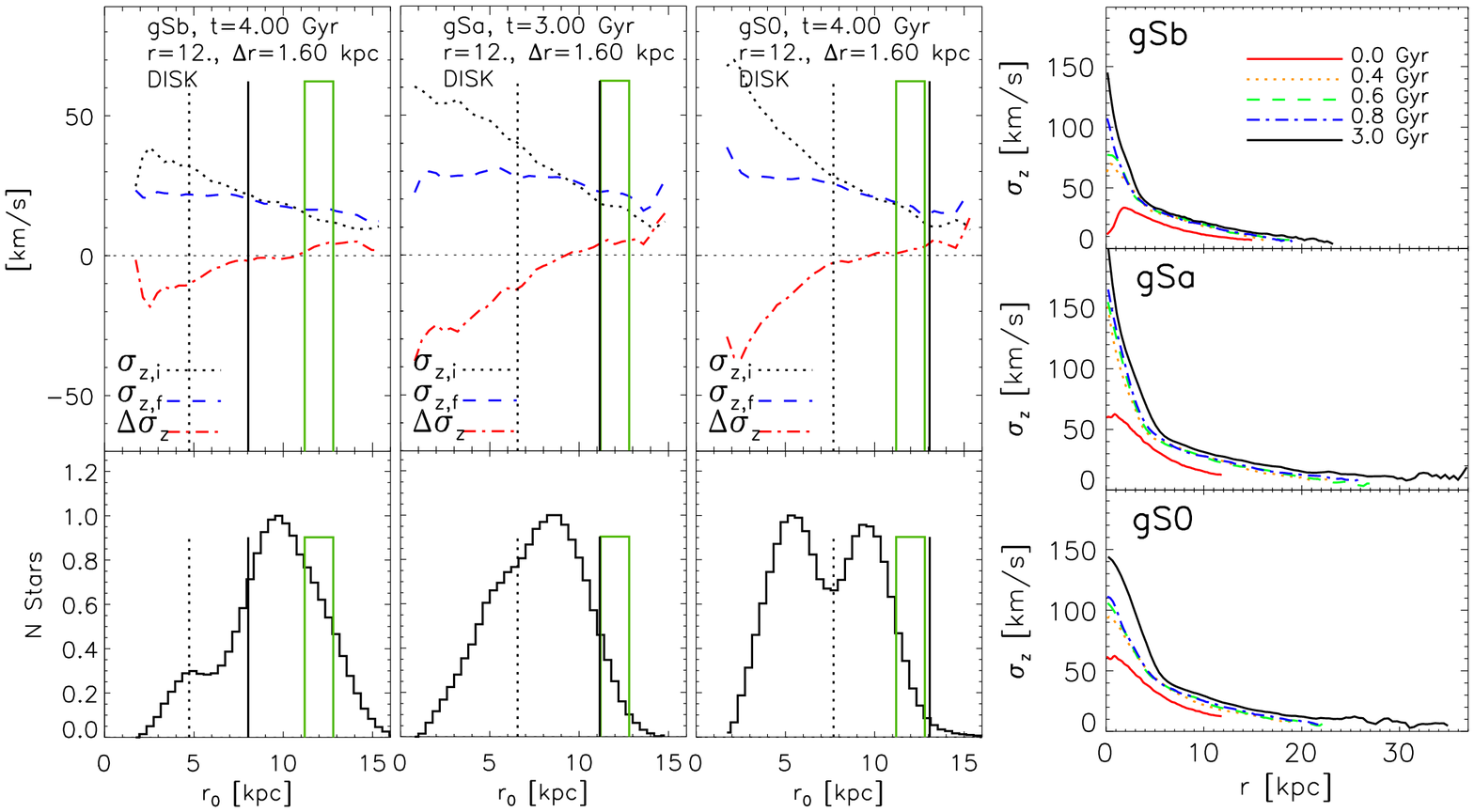} }
\caption{
{\bf Top three panels:} initial (dotted black), final (dashed blue), and net (dot-dashed red) vertical velocity dispersions, $\sigma_{z,i}$, $\sigma_{z,f}$, and $\Delta \sigma_z$, respectively, as functions of the {\it initial} radius for particles ending up in the green bin at the final time. {\bf Bottom three panels:} the corresponding initial radial distributions of stars ending up in the green bin. The bar's CR and OLR are shown by the vertical dotted and solid black lines. {\bf Fourth column:} time evolution of the vertical velocity dispersion profiles for the gSb (top), gSa (middle), and gS0 (bottom) discs. 
}
\label{fig:3} 
\end{figure}

Stellar samples in the inner galactic discs possess high velocity dispersions in all three components. It has been expected that by migrating radially outwards, stars retain their random energies and thus thicken the disc. If this were true, then why did we not see sufficient disc thickening even for the strong large-scale disc mixing seen in Fig.~\ref{fig:1}? To answer this question, we consider the samples of stars used to estimate the scale-heights in Fig.~\ref{fig:2}. 
In the top three panels of Fig.~\ref{fig:3} we plot the initial (dotted black), final (dashed blue), and net (dot-dashed red) vertical velocity dispersions, $\sigma_{z,i}$, $\sigma_{z,f}$, and $\Delta \sigma_z$, respectively, as functions of the {\it initial} radius for particles ending up in the green bin at the final time. It is immediately apparent that stellar samples arriving from the inner disc decrease their velocity dispersion by as much as 50\% when coming from the smallest radii, with the effect diminishing the smaller the radial distance traveled. It is also notable that the stars coming from radii exterior to the green bin increase their velocity dispersion. The negative slope in $\sigma_{z,f}$ indicates that the outward migrators indeed heat the outer disc, but not significantly. As seen in the initial radial distributions at the bottom three panels of Fig.~\ref{fig:3}, the majority of stars populating the annuli at 2.5 disc scale-lengths, $r_d$, (green bin) have arrived/migrated from the inner disc. 

Note that the fact that stellar samples migrating outwards decrease their vertical energy does not mean that the discs cool down. The fourth column of Fig.~\ref{fig:3} plots the time evolution of the vertical velocity dispersion profiles for gSb (top), gSa (middle), and gS0 (bottom). An increase in $\sigma_z$ is seen at every radius, as expected.

To understand this cooling process we consider a simple galactic disc model, where the vertical action of stars, $J_z$, is conserved as they shift guiding radii (i.e., migrate). This is true in the limit that (i) the vertical motion decouples from the planar motion (a good assumption when $z_{\max}\ll r$ which holds for most disc stars), and (ii) the stars move outwards much more slowly than their vertical oscillations (probably not really the case). We assume that for a given radius, $J_z = E_z/\nu=Const.$, where $E_z$ and $\nu$ are the vertical energy and frequency, respectively. From Gauss' law and Poisson's equations, $\nu\sim\sqrt{2 \pi G \Sigma}$, therefore, as stars migrate radially, their vertical frequency changes as $\nu(r)\sim\exp(-r/2r_d)$, given that the stellar density varies as $\Sigma\sim\exp(-r/r_d)$. To conserve the mean vertical action of a set of stars migrating outwards, we then require that their mean vertical energy, $\langle E_z\rangle\sim\sigma_z^2$, decreases as $\nu$. Therefore, the vertical velocity dispersion of a stellar sample migrating radially outwards will decrease as $\sigma_z(r)\sim\exp(-r/4r_d)$. On the other hand, the vertical velocity dispersion in the ambient (non migrating) disc component changes with radius as $\sigma_z(r)\sim\sqrt{\Sigma(r) h}$ (e.g., eq.~7 by \cite{vanderkruit11}). Assuming the scale-height, $h$, remains constant, we find that $\sigma_z(r)\sim\exp(-r/2r_d)$. Thus, while a star's vertical energy does decay adiabatically as it migrates outwards, it does not decay as fast as the underlying typical vertical energy of disc stars and, therefore, still increases the vertical temperature of the disc where it arrives. This is in agreement with our numerical results, where we find that stellar samples from the inner disc arrive at $2.5r_d$ with a velocity dispersion slightly larger than the typical values for that particular radius (top three panels of Fig.~\ref{fig:3}).

For the simple model considered, the same argument also holds for the epicyclic energy $E_p = \kappa J_p$, where the radial epicyclic frequency $\kappa\sim1/r$ (valid for a flat rotation curve), so that orbits conserving $J_p$ loose energy $E_p$. 
This would also suggest that the rates of change in $\sigma_z$ and $\sigma_r$ are different for the same sample of migrating stars: because in the outer disc at $r>r_d$, $\nu$ decreases faster than $\kappa$, $\sigma_r$ is less reduced. This is exactly what we find in plots similar to the upper three panels of Fig.~\ref{fig:3}, but for $\sigma_r$ (not shown): the outer disc is heated much more radially than vertically due to the migrated stars. Consequently, if one were to consider, for instance, an analytic model for radial migration in a 2D razor-thin disc using $J_p$ as a proxy for extrapolating the full distribution function (including the vertical direction), one might be fooled into thinking that a hot thick disc component can be created.

\section{Conclusions}

In this work we have used Tree-SPH N-body simulations to study the effect of internal disc evolution on disc thickening. All analyzed models were found to develop strong multiple patterns, resulting in very efficient radial disc mixing. Surprisingly, we found that discs with initial scale-height of $\sim300$~pc only double their thickness even for bar and spirals much stronger than those of the Milky Way. We showed that the reason for this is that stellar samples migrating from the inner disc decrease their vertical velocity dispersion. Therefore, although at any given radius the disc heats with time, the migrated stars do not contribute significantly to the disc thickness.  

Using a simple galactic disc model, where both the radial, $J_p=E_p/\kappa$, and vertical, $J_z=E_z/\nu$, actions, are conserved, we explained the decrease of velocity dispersion of outward migrators as adiabatic disc cooling. Due to the faster radial decrease of $\kappa$ compared to $\nu$, this model predicts that radial migration heats discs more radially than vertically, in agreement with the simulations (\cite{minchev12}). While stellar samples migrating outwards lose vertical energy, $\langle E_z\rangle\sim\sigma_z^2$, their reduced $\sigma_z$ is still higher than what is expected from extrapolating $\sigma_z\sim\Sigma(r)$. Thus, radial migration should lead to disc flaring. We indeed find that scale-height increases with radius in our simulations, however, not significantly so for $r<5$ scale-lengths. Stellar radial migration also mixes the radial distributions of the actions $\langle J_p\rangle$ and $\langle J_z\rangle$: the more the disc mixes, the weaker their variation with radius. Therefore, even if the migration rate remains constant, we can expect less heating at later times. 

Our simulations did not consider gas accretion. In a more realistic case, the mass increase due to continuous thin disc accretion would contract the thickened discs and further decrease the scale-heights \cite{bournaud09}. Therefore, what we find in this work may be an upper limit on the disc thickening caused by radial migration. Considering that the MW thick disc stars are older than 8~Gyr and that thick discs are observed at high redshifts, we expect that the 3-4~Gyr of time evolution we follow here is sufficient for the problem we address. In other words, if thick discs could be caused by radial migration, then that should occur early on, within the first 1/3 of a galaxy lifetime (even much faster if we include chemical considerations, e.g., attempt to explain the chemical shifts observed for different abundance rations between the thick and thin discs -- see \cite{chiappini09}). 

For the gSa and gS0 models, bulge contribution to the thick disc beyond $\sim$~2.5 disc scale-lengths is only about 1\%. For a MW-like bulge (the gSb model) no stars reach this radius (\cite{minchev12}). We conclude that, although radial migration does indeed contribute to the disc thickening, the help from an additional mechanism may be needed to account for thick disc formation, such as heating by mergers and/or accretion of satellites. At this time the only viable mechanism for purely secular thick disc formation remains the heating from turbulent, gas-rich clumpy discs at high redshift (\cite{bournaud09}).

\end{document}